\documentstyle[preprint,aps]{revtex}
\tighten
\begin{document}
\preprint{UAB--FT--371, gr-qc/9506045}
\draft

\title{Comment on ``Perturbative Method to solve fourth-order Gravity Field
Equations"}

\author{Manuela Campanelli,}
\address{Fakult\"at f\"ur Physik der Universit\"at Konstanz,
Postfach 5560 M 674, D - 78434 Konstanz, Germany,\\ and \\
IFAE - Grupo de F\'\i sica Te\'orica, Universidad Aut\'onoma de
Barcelona, E-08193 Bellaterra (Barcelona), Spain.}
\date{\today}
\maketitle

\begin{abstract}
We reconsider the cosmic string perturbative solution to the classical
fourth-order gravity field equations, obtained in Ref.\cite{CLA94}, and we
obtain that static, cylindricaly symmetric gauge cosmic strings,
with constant energy density, can contain only $\beta$-terms in the first
order corrections to the interior gravitational field, while the
exact exterior solution is a conical spacetime with deficit angle
$D=8\pi\mu$.
\end{abstract}


In Ref.\cite{CLA94} we considered the higher order derivative theory of
gravity derived from the action
\begin{eqnarray}
I=I_{G}+I_{m}={1\over 16\pi}\int d^4x\sqrt{-g}\bigg\{-2\Lambda + R
+ \alpha R^2 + \beta R_{\mu\nu}R^{\mu\nu} + 16\pi{\cal L}_m\bigg\}~~,
\label{QA}
\end{eqnarray}
(in units where $G=c=\hbar=1$)
where the coupling constants $\alpha$ and $\beta$ were supposed to be of
the order of the Planck length squared $l_{pl}^2$.
Moreover, as it is well known, the coupling constants $\alpha$ and $\beta$
must fulfill the no-tachyon constraints
\begin{equation}
3\alpha + \beta\geq 0~~~~,~~~~  \beta\leq 0~, \label{NTC}
\end{equation}
which can be deduced linearizing and asking for a real
mass for, both the scalar field $\phi$ related to $R$
and the spin-two field $\psi_{\mu\nu}$ related to $R_{\mu\nu}$
(see Ref. \cite{AEL93} for further details).

In Ref.\cite{CLA94}, we have developed a method to solve
the field equations of the quadratic gravitational theory in four
dimensions coupled to matter. The quadratic terms are written
as a function of the matter stress tensor and its derivatives in such
a way to have, order by order, Einsteinian field equations with
an effective $T_{\mu\nu}$ as source.
By successive perturbations around a solution to Einstein's gravity,
which for us represent the zeroth order, one can build up approximated
solutions.

For the perturbative approach to properly work we consider relatively
small curvatures, such that
\begin{equation}
\alpha |R|\ll 1 ~~~~,~~~~|\beta R_{\mu\nu}|\ll 1~, \label{SCC}
\end{equation}
According to our supposition for $\alpha$ and $\beta$ to be of
the order of the Planck length squared, this means that we deal with
underplanckian curvatures.

The field equations derived by extremizing the action \ (\ref{QA})
are
\begin{eqnarray}
R_{\mu\nu}-{1\over 2}Rg_{\mu\nu}+\Lambda g_{\mu\nu}+\alpha H_{\mu\nu}+
\beta I_{\mu\nu}=8\pi T_{\mu\nu}~,  \label{QFE}
\end{eqnarray}
where
\begin{mathletters}
\label{allQFE}
\begin{eqnarray}
H_{\mu\nu}=-2R_{;\mu\nu}+2g_{\mu\nu}\Box R-{1\over 2}g_{\mu\nu}R^2
+2R R_{\mu\nu}~,
\label{QFEa}
\end{eqnarray}
\begin{eqnarray}
&&I_{\mu\nu}=-2R_{\mu~;\nu\tau}^{\tau} +\Box R_{\mu\nu}+
{1\over 2}g_{\mu\nu}\Box R+2R_{\mu}^{~\tau}R_{\tau\nu}
-{1\over 2}g_{\mu\nu}R_{\tau\sigma}R^{\tau\sigma}~. \label{QFEb}
\end{eqnarray}
\end{mathletters}
These equations can be rewritten to the {\it n-th order} approximation
as
\begin{eqnarray}
R_{\mu\nu}^{(n)}-{1\over 2}R^{(n)}g^{(n)}_{\mu\nu}+\Lambda
g^{(n)}_{\mu\nu}=8\pi T_{\mu\nu}^{\text{eff}~(n)}~,
\label{NQFE}
\end{eqnarray}
where the {\it zeroth-order} corresponds to the ordinary Einstein
equations and  where
\begin{eqnarray}
T_{\mu\nu}^{\text{eff}~(n)}
=T_{\mu\nu}^{(n)}-{\alpha\over 8\pi} H^{(n-1)}_{\mu\nu}-{\beta\over 8\pi}
I^{(n-1)}_{\mu\nu}~,\label{EEMT}
\end{eqnarray}
acts as an effective energy-momentum tensor satisfying order by order the
conservation law.

In paper\cite{CLA94}, we applied this perturbative procedure to find
gauge cosmic string solutions up to first order in the coupling constant
$\alpha$ and $\beta$. For simplicity, we considered the case of an
infinite straigth
static gauge cosmic string of zero $r_0$ radius lying on the
$z$-axis and charaterized by an energy momentum tensor
(given in formula (17)\cite{CLA94}) proportional to a delta function.
However, from the considerations below it appears clear that the only
gravitational field outside the cosmic string allowed in this  model
is not that of Eqs. (22) and (23) of Ref.\cite{CLA94}, but it is
given by the metric expression\cite{V81,G85,L85},
\begin{equation}
ds^2=-dt^2+dz^2+dr^2+(1-4\mu)^2r^2d\phi^2 ~~,\label{EVSM}
\end{equation}
with coordinate ranges: $-\infty<t<+\infty$, $-\infty<z<+\infty$,
$0\leq\phi<2\pi$, $r_0(1-4\mu)^{-1}\sin\theta_M=r_b<r<+\infty$ (if
$\theta_M<\pi/2$), and the mass per unit length in the string is
$\mu={1\over 4}(1-\cos{\theta_M})$, when $0<\mu<1/4$.
In the weak field limit this geometry corresponds to that of a troncated
cone with deficit angle $D=2\pi(1-\cos{\theta_M})=8\pi\mu$.
As for this metric $R^{\tau}_{~\sigma\gamma\delta}=0$ so that
$R_{\mu}^{~\nu}=0$ and $R=0$, we have that $T_{\mu}^{~\nu}=0$.
Thus, this is an {\it exact} solution of the field equations \ (\ref{QFE}).
In fact, in general, vacuum solutions to Einstein equations
(even with cosmological constant) are solutions to the quadratic theory
(the converse, in general, is not true).

Indeed, the corrections due to the quadratic terms in the
gravitational action will only come in a cosmic string where
$T_{\mu\nu}\neq 0$.
Thus, the appropriate model of gauge cosmic strings to be considered,
in this case, is that of straight tubes, localized along the direction
of the $z$-axis, having a finite size radius $r_0\sim 1/\sqrt{\mu\lambda}$
($\lambda$ is a coupling constant of the tipical boson squared mass)
and the only non zero pressure component $P_z=-\rho$.
The most general expression of a static metric with cylindrical
symmetry  and Lorentz invariance in the $(t, z)$ plane, in cylindrical
coordinates ($0\leq\phi<2\pi$), reads
\begin{equation}
ds^2=A(r)(-dt^2+dz^2) + dr^2+r^2B(r)d\phi^2 ~. \label{GIM}
\end{equation}
The exact metric solution of Einstein equations,
in the case of constant energy density
$\rho(r)=\rho$, is given by\cite{G85}
\begin{equation}
ds^2=-dt^2+dz^2+dr^2+r_0^2\sin^2{(r/r_0)}d\phi^2 ~~,\label{EIM}
\end{equation}
For this metric the only non-zero Christoffel's
symbols are $\Gamma^{(0)r}_{~~\phi\phi}=-r_0\sin{(r/r_0)}\cos{(r/r_0)}$ and
$\Gamma_{~~r\phi}^{(0)\phi}=\Gamma_{~~\phi r}^{(0)\phi}=
r_0^{-1}\cot{(r/r_0)}$; the only non-zero components of the Ricci tensor are
$R_{~~r}^{(0)r}=R_{~~\phi}^{(0)\phi}=r_0^{-2}$; the Ricci scalar is
$R^{(0)}=2r_0^{-2}$, and the only non zero components of the energy-momentum
tensor are $T^{(0)t}_{~~t}=T^{(0)z}_{~~z}=-\rho=-(1/8\pi r_0^2)$.
Note that, since
$T^{(0)t}_{~~t}=-\rho$ and $T^{(0)z}_{~~z}=P_z$, the only pressure component
is exaclty $P_z=-\rho=-(1/8\pi r_0^2)$, and thus, in this
static model the string excerts no Newtonian attraction on a particle
that is at rest with respect to it.

To compute the first-order solutions
(in the coupling constants $\alpha$ and $\beta$) to the fourth order field
equations \ (\ref{QFE}), first we have to evaluate the first-order
effective energy-momentum tensor, given by
Eq. \ (\ref{EEMT}); this calculation is straighforward since
Eqs \ (\ref{allQFE}) greatly simplify when $T_{\mu\nu}$
is diagonal and its components depend essentially on only one coordinate,
(not sum over $\mu$)
\begin{mathletters}
\label{allQFER}
\begin{eqnarray}
&&{1\over 8\pi}H_{\mu\mu}^{(0)}=2\bigg\{T_{,rr}\delta^r_{\mu}-
\Gamma^r_{\mu\mu}T_{,r}-g_{\mu\mu}\bigg[g^{rr}T_{,rr}-
g^{\tau\tau}\Gamma^r_{\tau\tau}T_{,r}
-2\pi T^2 +8\pi TT_{\mu}^{~\mu} \bigg]\bigg\}~,
\label{QFERa}
\end{eqnarray}
and
\begin{eqnarray}
&&{1\over 8\pi}I_{\mu\mu}^{(0)}=T_{,rr}\delta^r_{\mu}-\Gamma^r_{\mu\mu}
T_{,r}\cr\cr
&&-2\bigg[\big(T_{r~,rr}^{r}+\Gamma^{\tau}_{r\tau}(T_{r~,r}^{r}-
T_{\tau~,r}^{\tau})\big)\delta^r_{\mu}-\Gamma^r_{\mu\mu}T_{r ~,r}^{r}
+(\Gamma^r_{\mu\mu ,r}-\Gamma^{\sigma}_{r\mu}\Gamma^r_{\mu\sigma})
(T_{\mu}^{~\mu}-T_{\mu}^{~\mu})
\bigg]\cr\cr
&&-g_{\mu\mu}\bigg[g^{rr}T_{,rr}-g^{rr}T_{\mu~,rr}^{~\mu}+
g^{\tau\tau}\Gamma^r_{\tau\tau}(T_{\mu ~,r}^{\mu}-T_{,r})
+4g^{rr}\Gamma^{\mu}_{r\mu}\Gamma^{\mu}_{\mu ,r}\cr\cr
&&-(4g^{\tau\tau}
(\Gamma^{\mu}_{\mu\tau})^2+2g^{\tau\tau}\Gamma^{\mu}_{\mu\rho}
\Gamma^{\rho}_{\tau\tau}-2g^{rr}\Gamma^{\mu}_{r\mu ,r})
T_{\mu}^{~\mu}+16\pi TT_{\mu}^{~\mu}-16\pi (T_{\mu}^{~\mu})^2
\bigg] ~.
\label{QFERb}
\end{eqnarray}
\end{mathletters}
Note that this last expression corrects Eq. (16) of Ref.\cite{CLA94}.
Metric and energy-momentum dependence is with respect to the zeroth order,
i.e. solution of usual Einstein's equations, Eq.\ (\ref{EIM}).
Thus, the components of the first order effective energy momentum are
\begin{eqnarray}
&&T^{(1){\rm eff}r}_{~~~r}=T^{(1){\rm eff}\phi}_{~~~\phi}=-32\pi
\alpha\rho^2\cr\cr
&&T^{(1){\rm eff}t}_{~~~t}=T^{(1){\rm eff}z}_{~~~z}=-\rho+16\pi
\beta\rho^2~.\label{TE}
\end{eqnarray}
The fundamental requirement that $T^{(1){\rm
eff}}_{\mu\nu}$ must be conserved in the whole spacetime,
and, thus, also on the boundary of the interior static solution
$r_{b_i}$, with $\rho(r)=\rho\theta(r_{b_i}-r)$, where $\theta$ is a step
function, give us the condition that the coupling constant $\alpha$
must vanish.
The question of finding solutions, when the energy density is not
constant but a generic function $\rho(r)$, is not such a
straighforward problem to solve, thus, the
possibility of a non vanishing coupling constant, $\alpha>0$, remains open.

Now, we plug metric \ (\ref{GIM}),
into Eqs. \ (\ref{NQFE}) and \ (\ref{EEMT}). The interior solution
in order to satisfy the boundary conditions,
at $r=r_b$ must match the exterior metric given by Eq. \ (\ref{EVSM}).
Using the criterion of Israel\cite{I66}, we impose the following
jump conditions:
$g_{\mu\nu}\big|^{ext}_{r_{b_e}}=g_{\mu\nu}\big|^{int}_{r_{b_i}}$ and
$\partial_rg_{\mu\nu}\big|^{ext}_{r_{b_e}}
=\partial_rg_{\mu\nu}\big|^{int}_{r_{b_i}}$,
i. e. the metrics and its derivatives should match as the boundary is
approached from each side. From the exterior side the boundary is
at the coordinate
$r_{b_e}=r_0(1-4\mu)^{-1}\sin{\theta_M}=r_0\tan{\theta_M}$ and from the
interior side at the coordinate $r_{b_i}=r_0\theta_M$.
Finally, for the first order metric given  we get
\begin{mathletters}
\label{allFIM}
\begin{eqnarray}
&&A^{(1)}(r)=1\cr\cr
\label{FIMa}
&&B^{(1)}(r)={r_0^2\over r^2}\sin^2{\left({r\over r_0}\right)}
+2{\beta\over r^2}\sin^2{\left({r\over r_0}\right)}
\left[\theta_M-{r\over r_0}\cot\theta_M\right]
\label{FIMb}
\end{eqnarray}
\end{mathletters}
where parameters are $r_{b_i}=r_0\theta_M$ and
$\mu={1\over 4}(1-\cos\theta_M)$.
Moreover, following Linet\cite{L85} it is straighforward to verify that
this metric is perfectly finite and regular at $r=0$.

In agreement with some previous results (Ref.\cite{AEL93}),
it appears clear that the $\beta$-terms corrections to the string interior
metric could only modify the dynamics of eventual
collisions of cosmic strings which involve very short range
interactions. As we considered $1<\mu<1/4$ ($0<\theta_M<\pi/2$),
the corrections here appear to give a negative contribution to
$B(r)$ of \ (\ref{GIM}), when
$r_0\theta_M<r<r_0\theta_M\tan\theta_M$, and positive
contribution, when $r>r_0\theta_M\tan\theta_M$,
since $\beta\leq 0$ from the no-tachyons constraints \ (\ref{NTC}).

The outcome of future numerical simulations for collisions of
cosmic strings confronted with observations may allow to put some
constraints on the coupling constants $\alpha$ and $\beta$.
Long-range corrections to the dynamics of structure formation scenarios
are absent since it is the exterior gravitational field who play the
important role in this case, and it is exaclty the same in both,
General relativity and higher order gravity.

\begin{acknowledgments}
The author would like to thank J. Audretsch, A. Campos, J. Garriga and
C. O. Lousto for helpful conversations.
This work was partially supported by the Directorate General for
Science, Research and Development of the Commission of the European
Community and CICYT AEN 93-0474. M.C. holds an
scholarship from the Deutscher Akademischer Austauschdienst.
\end{acknowledgments}

\end{document}